\documentclass[graybox, envcountchap, authoryear, natbib]{svmult}
\usepackage{mathptmx}       
\usepackage{amsmath} 
\usepackage{amsbsy}
\usepackage{amssymb}
\usepackage{mathrsfs} 
\usepackage{helvet}         
\usepackage{courier}        
\usepackage{type1cm}        
\usepackage{verbatim}
\usepackage{natbib}
\usepackage{sidecap} 

\usepackage{makeidx}         
\usepackage{graphicx}        
\usepackage{multicol}        
\usepackage[bottom]{footmisc}

\makeindex             

\newcommand{\apj}{ApJ} 
\newcommand{\aap}{A\&A}

\newcommand{\apjl}{ApJL}

\newcommand{\mnras}{MNRAS}
\newcommand{\aj}{AJ}
\newcommand{\apjs}{ApJS}
\newcommand{\nat}{{\it Nature}}
\newcommand{\araa}{ARA\&A}

\begin{document}

\title*{Observational Diagnostics of Gas Flows: Insights from Cosmological Simulations}
\titlerunning{Simulations of Galactic Accretion Diagnostics}
\author{Claude-Andr\'e Faucher-Gigu\`ere}
\institute{Department of Physics and Astronomy and Center for Interdisciplinary Exploration and Research in Astrophysics (CIERA), Northwestern University, 2145 Sheridan Road, Evanston, IL 60208, USA. cgiguere@northwestern.edu.}
\maketitle

\abstract{Galactic accretion interacts in complex ways with gaseous halos, including galactic winds. As a result, observational diagnostics typically probe a range of intertwined physical phenomena. 
Because of this complexity, cosmological hydrodynamic simulations have played a key role in developing observational diagnostics of galactic accretion. 
In this chapter, we review the status of different observational diagnostics of circumgalactic gas flows, in both absorption (galaxy pair and down-the-barrel observations in neutral hydrogen and metals; kinematic and azimuthal angle diagnostics; the cosmological column density distribution; and metallicity) and emission (Ly$\alpha$; UV metal lines; and diffuse X-rays). 
We conclude that there is no simple and robust way to identify galactic accretion in individual measurements. 
Rather, progress in testing galactic accretion models is likely to come from systematic, statistical comparisons of simulation predictions with observations. 
We discuss specific areas where progress is likely to be particularly fruitful over the next few years.}

\section{Introduction}
\label{sec:1}
In recent years, there has been a growing realization that the ``cosmic baryon cycle'' is both a primary driver and a primary regulator of galaxy formation. 
Continuous accretion of gas from the intergalactic medium (IGM) is necessary to sustain observed star formation rates (SFRs) over a Hubble time \citep[e.g.,][]{2008ApJ...674..151E, 2009ApJ...696.1543P, 2010ApJ...717..323B}. 
However, models in which the intergalactic gas accreted by galaxies is efficiently converted into stars produce galaxies with stellar masses that exceed observed ones by an order of magnitude or more \citep[e.g.,][]{1991ApJ...379...52W, 1995MNRAS.275...56N, 2009MNRAS.396.2332K}. 
In the latest generation of models, star formation-driven galactic winds regulate galaxy growth below $\sim L^{\star}$ by ejecting back into the IGM most of the accreted gas before is has time to turn into stars \citep[see the review by][]{2015ARA&A..53...51S}. 
Despite a broad consensus regarding the importance of inflows and outflows in galaxy evolution, many questions regarding their nature and effects remain at the forefront of current research. 

For example, many cosmological simulations and semi-analytic models now suggest that wind recycling (the fallback of gas previously ejected in galactic winds) plays an important role in shaping the galaxy stellar mass function and setting the level of late-time galactic accretion \citep{2010MNRAS.406.2325O, 2013MNRAS.431.3373H, Angles-16}. 
While galactic accretion is a generic prediction of cosmological simulations \citep[e.g.,][]{2005MNRAS.363....2K, 2009MNRAS.395..160K, 2009Natur.457..451D, 2009ApJ...694..396B, 2011MNRAS.417.2982F, 2011MNRAS.414.2458V}, its properties are subject to uncertainties in how the accretion flows are affected by shocks and hydrodynamical instabilities as they interact with galaxy halos \citep[e.g.,][]{2003MNRAS.345..349B, 2013MNRAS.429.3353N, 2016arXiv160606289M}. 
Galactic winds are driven by feedback processes that operate on the scale of individual star-forming regions, which are generally not well resolved in current simulations. 
As a result, detailed properties such as their phase structure remain uncertain even in today's highest resolution zoom-in simulations of galaxy formation \citep[e.g.,][]{2013ApJ...765...89S, 2014MNRAS.445..581H, 2014MNRAS.442.3745M, 2015ApJ...804...18A}. 
In large-volume cosmological simulations, it is not yet possible to resolve how galactic winds are launched so even the bulk properties of galactic winds in such simulations are typically tuned to match observables such as the galaxy stellar mass function \citep[e.g.,][]{2011MNRAS.415...11D, 2014MNRAS.444.1518V, 2015MNRAS.446..521S}. 
Theoretical predictions for inflows and outflows are furthermore complicated by the fact that inflows and outflows inevitably interact with each other \citep[e.g.,][]{2011MNRAS.414.2458V, 2011MNRAS.417.2982F, 2015MNRAS.449..987F, 2015MNRAS.448...59N}.

The importance of inflows and outflows for galaxy evolution, as well as the significant theoretical uncertainties, imply that observations of these processes are critical to test and inform galaxy formation theories. 
Since observational techniques for probing inflows and outflows generally provide only fragmentary information about the physical nature of the observed gas (e.g., 1D skewers through galactic halos for typical quasar absorption line measurements), forward modeling using cosmological simulations and comparing the simulations with observations will likely continue to play a central role in disentangling these processes. 
In this chapter, we review the current status of using cosmological simulations to develop observational diagnostics of galactic accretion. 
Since the dynamics inflows and outflows are intertwined in the circum-galactic medium (CGM), this chapter will also cover relevant outflow diagnostics.  

This chapter is largely organized around our group's research on the topic, but attempts to provide a broad review of theoretical research relevant to interpreting recent and upcoming observations. 
The chapter is divided into two main sections, one on absorption diagnostics (\S \ref{sec:absorption}) and one on emission diagnostics (\S \ref{sec:emission}). 
Interspersed within our discussion of different observational diagnostics, we include some remarks on numerical uncertainties and the sensitivity of different predictions to the numerical method employed. 
We conclude in \S \ref{sec:conclusions} with a synthesis of lessons from existing simulations of galactic accretion and comparisons with observations, and suggest some promising directions for future work. 
We focus on observational diagnostics applicable to galaxies other than the Milky Way. 

\begin{figure}
\sidecaption
\includegraphics[scale=.525]{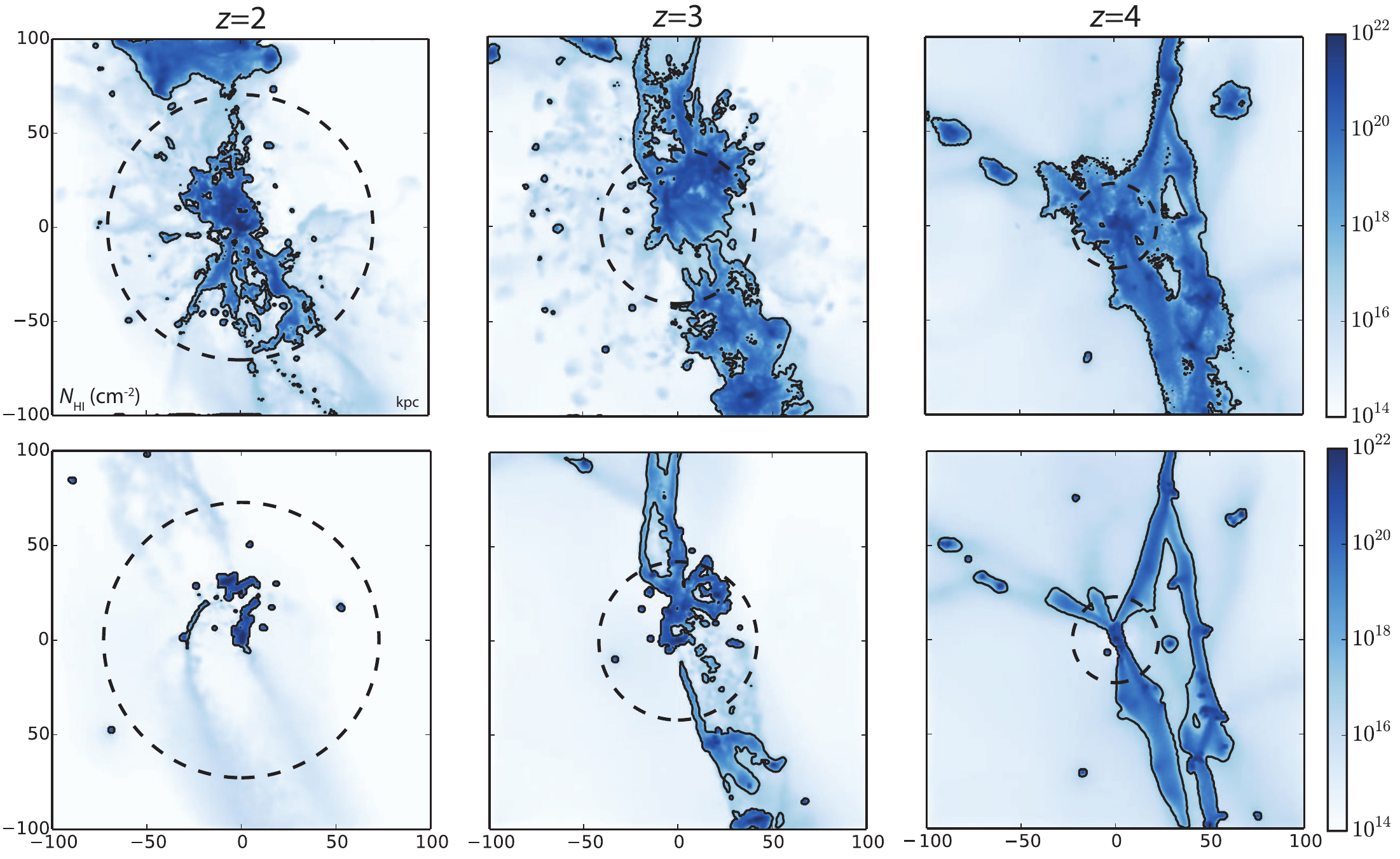}
\caption{\emph{Top:} H I maps for a low-mass LBG simulation with stellar feedback from the FIRE project at $z=2,~3,~{\rm and}~4$ ($M_{\rm  h}(z=2)=3\times10^{11}$ M$_{\odot}$).
\emph{Bottom:} Simulation from the same initial conditions but without galactic winds. 
The virial radius of the halo is indicated in each panel by the dashed circles and 
Lyman limit systems are indicated by solid contours. 
Stellar feedback increases the 
covering fractions in galaxy halos both by directly ejecting cool gas from galaxies and through the interaction of galactic winds with cosmological inflows. 
At $z=2$, LLSs in this example are almost exclusively restricted to galaxies and their immediate vicinity absent galactic winds. Length scales are consistent across rows and columns. Adapted from \cite{2015MNRAS.449..987F}.}
\label{fig:fbk_HI_dep}
\end{figure}

\section{Absorption Diagnostics}
\label{sec:absorption}
We divide our discussion of absorption diagnostics into different observational statistics.

\subsection{H I Covering Fractions}
\label{sec:HI_coverings}
Covering fractions of absorbers within different impact parameters from foreground galaxies have been extensively modeled using simulations and provide the most basic consistency test between simulations and observations.
Over the last decade, large observational datasets on absorption by the CGM gas flows has been assembled using quasar spectra transverse to galaxies of different types and at different redshifts. 
For example, this technique has been applied at both low and high redshifts to foreground dwarf galaxies \citep[e.g.,][]{2014ApJ...796..136B}, damped Ly$\alpha$ absorbers \citep[e.g.,][]{2015ApJ...808...38R}, luminous red galaxies \citep[e.g.,][]{2010ApJ...716.1263G}, $\sim L^{\star}$ star-forming galaxies \citep[e.g.,][]{2003ApJ...584...45A, 2011Sci...334..948T, 2012ApJ...750...67R, 2014MNRAS.445..794T}, and quasars \citep[e.g.,][]{2006ApJ...651...61H, 2013ApJ...762L..19P}. 
In a study of $z\sim2-3$ Lyman break galaxies (LBGs), \cite{2010ApJ...717..289S} also demonstrated that useful constraints on the CGM can be extracted from spectra of ordinary, fainter background galaxies \citep[see also][]{2011ApJ...743...10B}. 
With the advent of 30-m class ground-based telescopes in the next decade, spectroscopy of background galaxies will become increasingly powerful as it becomes generically possible to obtain spectra of multiple sight lines through the halos of individual foreground galaxies.

Lyman limit systems (LLSs; usually quantitatively defined as systems with HI column density $N_{\rm HI} \geq 10^{17.2}$ cm$^{-2}$) in particular are useful tracers of inflows and outflows, being dense enough to be closely associated with galaxy halos but not sufficiently dense to arise only in galactic disks.\footnote{Galactic disks are better traced by damped Ly$\alpha$ absorbers (DLAs; $N_{\rm HI}\geq 2\times10^{20}$ cm$^{-2}$; e.g., Wolfe et al. 2005; Neeleman et al. 2015\nocite{2005ARA&A..43..861W, 2015ApJ...800....7N}). 
At very high redshift, the increased cosmic mean density and declining cosmic ultraviolet background (UVB) cause absorbers of fixed HI column to probe structures more closely associated with the low-density IGM with increasing redshift \citep[e.g.,][]{2011ApJ...743...82M}. 
As a result, LLSs become associated with structures such as intergalactic filaments and some DLAs may arise in the CGM. 
There rapid increase in LLS incidence observed at $z\gtrsim3.5$ suggests that LLSs commonly arise outside galaxy halos at these redshifts \citep[][]{2013ApJ...775...78F} while the rapid evolution of the DLA metallicity distribution at $z\gtrsim5$ suggest that DLAs at these redshifts commonly arise outside galaxies \citep[][]{2014ApJ...782L..29R}.
}  
Cosmological simulations show that LLSs are good tracers of cool filamentary accretion, especially at high redshift ($z\sim2-4$) where these are most prevalent \citep[e.g.,][]{2011MNRAS.418.1796F, 2011MNRAS.412L.118F, 2011MNRAS.413L..51K, 2012MNRAS.424.2292G, 2013ApJ...765...89S, 2014ApJ...780...74F, 2015MNRAS.449..987F}. 

Radiative transfer is important to properly model LLSs since these systems are optically thick at the Lyman limit by definition. 
However, until recently most cosmological simulations computed ionization balance assuming that all systems are optically thin. 
In early studies using simple approximations for the ionization state of the gas, (e.g., Dekel et al. 2009\nocite{2009Natur.457..451D} though see Kimm et al. 2011\nocite{2011MNRAS.413L..51K}), the predicted covering fractions of cold accretion streams were well in excess of observational constraints \citep[][]{2010ApJ...717..289S}. 
Properly processing simulations with ionizing radiative transfer -- thus allowing more accurate identification of strong HI absorbers -- showed that the LLS covering fractions of cold accretion streams are in fact quite small in simulations. 
As a result, the predicted small LLS covering fractions of cold accretion streams are consistent with present observational constraints \citep[$\lesssim 10\%$ for LBG-mass halos at $z\sim2$;][]{2011MNRAS.412L.118F, 2011MNRAS.418.1796F}. 

The more accurate treatments of radiative transfer actually revealed tension in the opposite direction. 
In an analysis of high-resolution quasar spectra transverse to $z\sim2-2.5$ LBGs, \cite{2012ApJ...750...67R} measured an LLS covering fraction within a projected virial radius of $30\pm14$\%, at face value a factor $\sim 3$ higher than cosmological simulations without strong galactic winds \citep[][]{2011MNRAS.412L.118F, 2011MNRAS.418.1796F}. 
This discrepancy has been plausibly resolved in the latest generation of cosmological simulations with stronger stellar feedback, necessary to produce realistic galaxy stellar masses \citep[][]{2014ApJ...780...74F, 2015MNRAS.449..987F}. 
These simulations showed that $\gtrsim50\%$ of the cool halo gas giving rise to LLSs around $z\sim2-3$ LBGs arises not from IGM accretion but rather from galactic winds. 
Figure \ref{fig:fbk_HI_dep} shows two simulations of the same low-mass LBG halo, one with strong galactic winds and one without galactic winds, at $z=2-4$. 

Galactic winds enhance LLS covering fractions in the simulations in two ways: 1) they eject cool interstellar gas into the CGM, and 2) they increase the cross section of inflows through hydrodynamic interactions. 
Importantly, it is not only the galactic wind from the central galaxy that interacts with infalling gas, but also outflows from nearby satellites. 
The latter effect is enhanced because satellites tend to be embedded in surrounding large-scale structure filaments. 
These filaments are ``puffed up'' by galactic winds from embedded galaxies. 
It is apparent from the example in Figure \ref{fig:fbk_HI_dep} that, absent galactic winds, the $z\sim2$ LLS covering fraction from accreting gas is very small and almost entirely limited to galaxies and their immediate vicinity (at higher redshifts, where the halos are less massive, the filamentary inflows give rise to more extended LLSs). 
This is because cool filamentary inflows tend to disappear in higher mass, lower redshift halos \citep[e.g.,][]{2005MNRAS.363....2K, 2006MNRAS.368....2D}. 
Most latest-generation cosmological simulations, including those from the EAGLE \citep[][]{2015MNRAS.446..521S}, Illustris \citep[][]{2014MNRAS.444.1518V}, and FIRE \citep[][]{2014MNRAS.445..581H}\footnote{See the FIRE project web site at http://fire.northwestern.edu.} projects, implement on-the-fly approximations based on local gas properties for self-shielding based on post-processing radiative transfer calculations \citep[][]{2010ApJ...725..633F, 2011MNRAS.418.1796F, 2013MNRAS.430.2427R}. 
\begin{figure}
\sidecaption
\includegraphics[scale=.75]{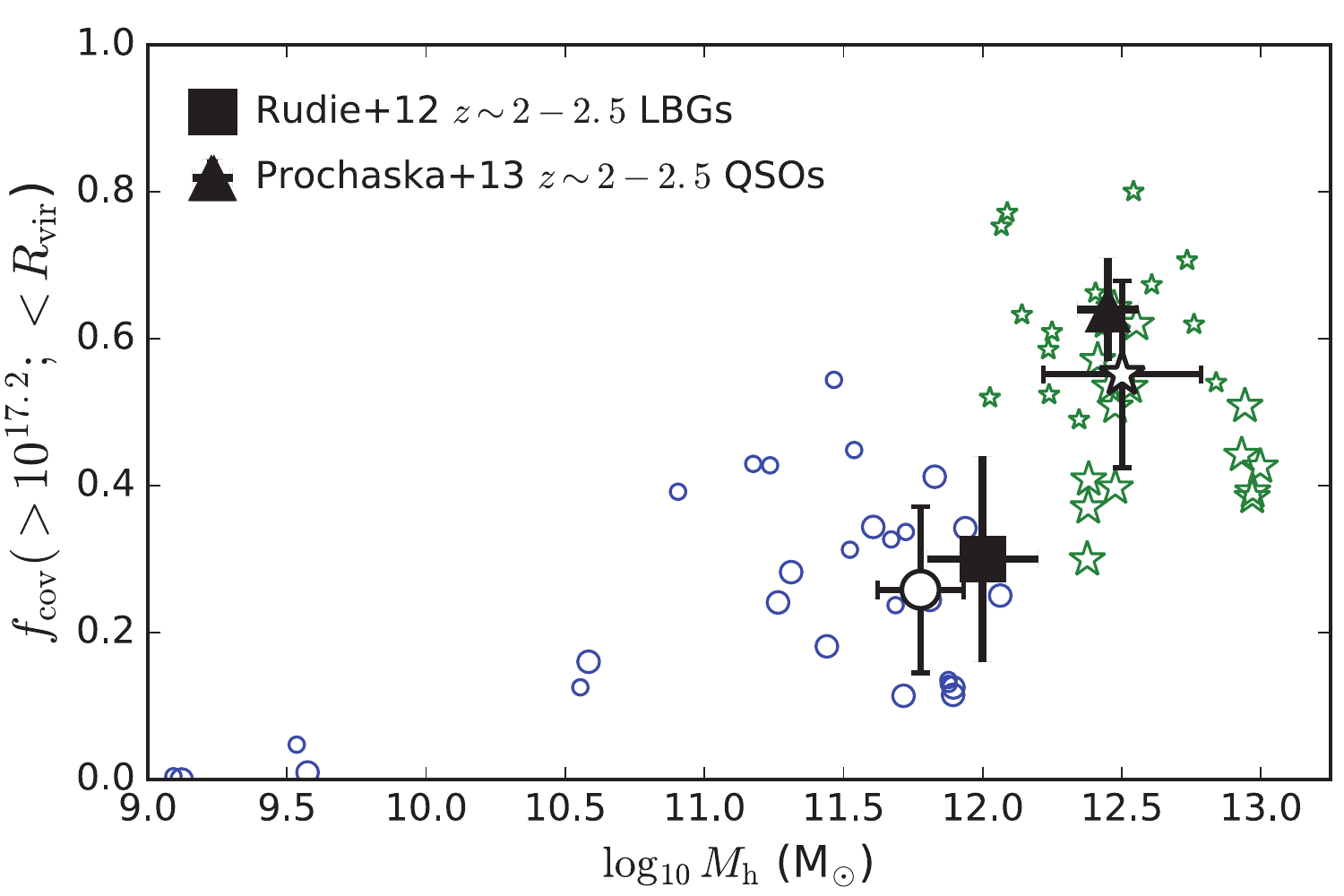}
\caption{\emph{Blue circles}: Lyman limit system (LLS) covering fractions within a projected virial radius for the high-resolution simulated halos from the FIRE project analyzed in \cite{2015MNRAS.449..987F}. 
For each simulated halo, covering fractions for 25 snapshots over the redshift interval $z=2-2.5$ are shown. The simulations are in good agreement with LLS covering fractions measured around LBGs in that redshift interval by \cite{2012ApJ...750...67R} (black square).
\emph{Green stars:} Covering fractions at $z=2$ (large) and $z=2.5$ (small) for the quasar-mass halos analyzed in \cite{2016MNRAS.461L..32F}. 
The quasar-mass simulated halos are compared to LLS measurements transverse to luminous quasars at $z\sim2-2.5$ by \cite{2013ApJ...762L..19P} (black triangle). 
The open black symbols show averages over simulated LBG-mass halos and QSO-mass halos, with the error bars showing the standard deviations of the simulated data points included in the averages. Figure from \cite{2016MNRAS.461L..32F}.}
\label{fig:fg16_summary}
\end{figure}

A current puzzle are the order unity LLS covering fractions measured in the halos of luminous quasars $z\sim2-2.5$. 
\citet{2013ApJ...762L..19P} reported an LLS covering fraction $f_{\rm cov}(>10^{17.2};~<R_{\rm vir})\approx0.64^{+0.06}_{-0.07}$ within a projected virial radius of  $z\sim2-2.5$ quasars \citep[see also][]{2014ApJ...796..140P}. 
This high LLS covering fraction should be compared to the lower fraction $f_{\rm cov}(10^{17.2};~<R_{\rm vir})=0.30\pm0.14$ measured by \cite{2012ApJ...750...67R} around $z\sim2-2.5$ Lyman break galaxies (LBGs) in the Keck Baryonic Structure Survey (KBSS). 
The LBGs in KBSS reside in dark matter halos of characteristic mass $M_{\rm h}\approx10^{12}$ M$_{\odot}$ \citep[][]{2005ApJ...620L..75A, 2012ApJ...752...39T}, a factor just $\sim3\times$ lower than quasars. 
Using cosmological zoom-in simulations with stellar feedback but neglecting AGN feedback, \citet[][]{2014ApJ...780...74F} and \citet[][]{2015MNRAS.449..987F} found simulated LLS covering fractions consistent with those measured in LBG halos (see also Shen et al. 2013\nocite{2013ApJ...765...89S}). 
In both studies, however, the most massive simulated halos failed to explain the LLS covering fraction measured around quasars by a large factor, suggesting that the presence of a luminous AGN could affect the properties of CGM gas on $\sim 100$ kpc scales. 

More recent simulations by \cite{2015MNRAS.452.2034R} and \cite{2016MNRAS.461L..32F} were able to match the covering fractions observed by \cite{2013ApJ...762L..19P} in quasar halos, but for different reasons. 
Recognizing that the distribution of halo masses probed by quasars is only crudely constrained by clustering measurements, \cite{2015MNRAS.452.2034R} made the optimistic assumption that all of \cite{2013ApJ...762L..19P}'s quasars are hosted in halos of mass \emph{greater than} the characteristic clustering mass $M_{\rm h}\approx 3\times10^{12}$ M$_{\odot}$. 
They then compared the quasar observations with the halos in the EAGLE simulation with mass above this threshold as a function of impact parameter in absolute units of proper distance. 
As a result, many of \cite{2015MNRAS.452.2034R}'s simulated LLSs are located at a smaller \emph{fraction} of the virial radius  than would be inferred assuming a constant virial radius corresponding to the characteristic quasar clustering halo mass (the assumption made in Prochaska et al. 2013). 
Since covering fractions decreases with increasing impact parameter, \cite{2015MNRAS.452.2034R}'s approach tends to boost the covering fractions, enough to bring them in agreement with those observed around real quasars. 
\cite{2015MNRAS.452.2034R}'s fiducial simulation included AGN feedback, but AGN feedback does not appear to play a significant role in explaining their results. 
\cite{2016MNRAS.461L..32F}'s simulations, on the other hand, are consistent with the $f_{\rm cov}(>10^{17.2};~<R_{\rm vir})$ value reported by \cite{2013ApJ...762L..19P}. 
\cite{2016MNRAS.461L..32F}'s simulations, from the FIRE project, included strong stellar feedback but no AGN feedback. 
Relative to the analysis of \cite{2015MNRAS.449..987F}, which focused on LBG-mass halos, \cite{2016MNRAS.461L..32F} analyzed a much larger set of quasar-mass halos (15 vs. 1) and the new halos were simulated at order-of-magnitude better mass resolution than the previous quasar-mass halo. 
\begin{figure}
\sidecaption
\includegraphics[scale=.315]{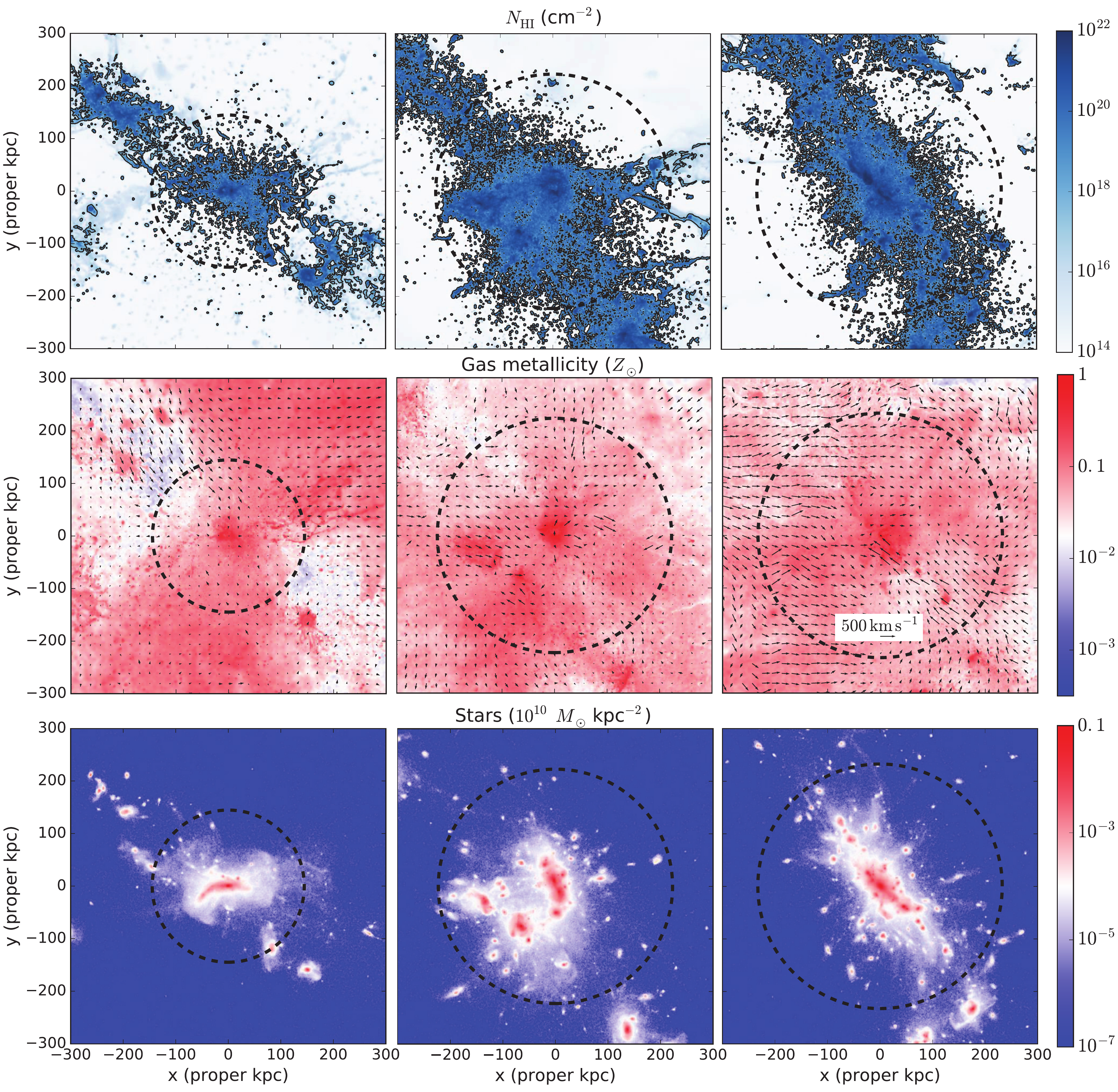}
\caption{
H I column density (top), gas-phase metallicity (middle) and stellar mass surface density (bottom) maps for three representative quasar-mass halos from the FIRE project at $z=2$ (from left to right: $M_{\rm h}(z=2)=(2.4, 8.8, 9.9)\times10^{12}$ M$_{\odot}$). 
The virial radius is indicated by dashed circle in each panel and solid contours indicate Lyman limit systems. 
The vectors on metallicity maps indicate projected mass-weighted velocities. 
The large-scale distribution of LLS gas correlates with the spatial distribution of satellite galaxies, indicating the importance of stellar feedback from satellites in producing large HI covering fractions. 
The velocity maps also show that the LLS structures with embedded satellites are typically infalling. 
Figure from \cite{2016MNRAS.461L..32F}.}
\label{fig:fg16_images} 
\end{figure}

Figure \ref{fig:fg16_summary} summarizes how the simulated LLS covering fractions of \cite{2015MNRAS.449..987F} and \cite{2016MNRAS.461L..32F} compare with observed covering fractions at $z\sim2-2.5$. For both LBG and quasar halos, the simulations rely on star formation-driven galactic winds to explain observations. 
\cite{2016MNRAS.461L..32F} performed a resolution convergence study of the covering fractions in quasar-mass halos, and found two important results. 
The first is that LLS covering fractions increase with increasing resolution. 
This is the primary reason why \cite{2015MNRAS.449..987F}'s earlier quasar-mass run fell short of reproducing observations. 
The second is that much of the LLS material in quasar-mass halos is due to galactic winds not from central galaxies but instead from lower-mass satellite galaxies. 
This is illustrated in Figure \ref{fig:fg16_images}, which shows that the spatial distribution of LLSs in quasar-mass halos correlates strongly with the spatial distribution of satellites. 
As in lower-mass halos, galactic winds from satellites both eject gas into the CGM and increase the cross section of large-scale structure filaments. 
The velocity maps in Figure \ref{fig:fg16_images} show that the LLS structures with embedded satellites are typically infalling, so these LLS are connected to galactic accretion, albeit somewhat indirectly. 

The gas particle mass in \cite{2016MNRAS.461L..32F}'s simulations of quasar-mass halos is $m_{\rm b}\approx3\times10^{4}$ M$_{\odot}$. 
At this resolution, Figure \ref{fig:fg16_summary} shows that the simulations are in good agreement with observations of massive halos. 
Since these covering fractions are not far from unity, they are necessarily saturating with increasing resolution. 
However, a comparison with lower-resolution simulations indicates that the CGM properties of quasar-mass halos may not be fully converged even in the highest-resolution simulations presently available. 
The stringent resolution requirements are in part due to the necessity of resolving the generation of galactic winds from satellites. 
Since the high simulated covering fractions in \cite{2016arXiv160700016F} do not require AGN feedback, one prediction is that similarly massive halos without a quasar -- such as may be selected based on high stellar mass or SFR - should show similarly high covering fractions. 
In an analysis of three sight lines with impact parameter $\sim100-200$ proper kpc from $z\sim2-2.5$ sub-millimeter galaxies (SMGs), \cite{2016arXiv160700016F} did not find compelling evidence for LLS-strength absorbers. 
If the simulations are correct and this observational result persists when the SMG sample is increased, it would suggest that some SMGs are hosted by halos significantly less massive than luminous quasars at $z\sim2$. 
A larger observational sample is, however, clearly needed to firm up the statistical significance of this observational result.

The ability to develop robust observational diagnostics of galactic accretion depends critically on the ability of numerical codes to properly capture the hydrodynamics of gas accretion, so we briefly digress to comment on this issue. 
In the example shown in Figure \ref{fig:fbk_HI_dep}, cold streams (as traces by LLSs) disappear in a slightly lower-mass halo ($M_{\rm  h}(z=2)=3\times10^{11}$ M$_{\odot}$) than predicted by some previous simulations \citep[see, e.g., the simulations similar-mass halos in][]{2011MNRAS.412L.118F} because older smooth particle hydrodynamics (SPH) simulations underestimated the destructive effects of shocks and hydrodynamical instabilities \citep[][]{2007MNRAS.380..963A, 2012MNRAS.424.2999S, 2013MNRAS.429.3353N}. 
Recent improvements to SPH algorithms \citep[e.g.,][]{2012MNRAS.422.3037R, 2013ApJ...768...44S, 2013MNRAS.428.2840H, 2014MNRAS.443.1173H} have greatly reduced the major historical differences with respect to grid-based codes, particular for fluid mixing instabilities. 
Overall, cold streams falling into galaxy halos tend to be more rapidly disrupted by hydrodynamic interactions with halo gas in codes of improved accuracy. 
The morphological differences in cold gas properties between hydrodynamic solvers are largest around the halo mass $M_{\rm h}\sim3\times10^{11}$ M$_{\odot}$ above which quasi-static hot atmospheres start to develop.    
In lower-mass halos, cold streams are generically present in halos simulated using different numerical methods, at least at $z=2$, which has been the subject of the most detailed simulation analyses. 
Interestingly, the most important differences overall for gas accretion between older SPH codes and grid-based codes are in the amount of ``hot mode'' accretion, i.e. the amount of hot gas that cools from hot atmospheres. 
Hot accretion is significantly more efficient in grid codes and updated SPH codes because spurious heating from the dissipation of turbulent energy on large scales prevents the same behavior in traditional SPH codes \citep[][]{2013MNRAS.429.3353N}. 
It is important to note, though, that most of the relevant direct code comparisons were performed on simplified cosmological simulations without strong galactic feedback. 
Observations of galaxy clusters clearly show that intra-cluster gas must be heated (likely by AGN feedback) to prevent a cooling catastrophe and avoid SFRs order-of-magnitude in excess of those observed in brightest cluster galaxies \citep[e.g.,][]{2007ARA&A..45..117M, 2012ARA&A..50..455F}. 
This heating suppresses hot mode accretion. 
It is not yet clear how much different numerical methods for hot mode accretion diverge when realistic feedback is included. 
Comparing the predictions of simulations with observations will continue to play a critical role in identifying limitations of the simulations. 

Before closing this section, we note that observations provide significantly more detail on the distribution of neutral hydrogen in galaxy halos than captured by the LLS covering fractions emphasized above, including better statistics on the incidence of (more numerous) lower-column absorbers and their line-of-sight velocity distributions \citep[e.g.,][]{2012ApJ...750...67R}. 
More comprehensive comparisons with simulations will be necessary to fully exploit the discriminating power of these observations for galactic inflow and outflow models.

\subsection{Metal Absorption Systems Transverse to Galaxies}
\label{sec:metal_abs}
Metal absorption is commonly observed out to $\sim0.5-1$ R$_{\rm vir}$ transverse to foreground galaxies of different types \citep[e.g.,][]{2005ApJ...629..636A, 2010ApJ...717..289S, 2012MNRAS.427.1238C, 2013ApJS..204...17W, 2014ApJ...796..136B, 2014MNRAS.445.2061L, 2015arXiv151006018L}. 
However, gas that is first accreting from the IGM is expected to be relatively metal-poor \citep[e.g.,][]{2012MNRAS.423.2991V}. 
In cosmological simulations with relatively weak stellar feedback, \cite{2011MNRAS.418.1796F} found that the mean metallicity of cold streams in $M_{\rm h}\sim10^{10}-10^{12}$ M$_{\odot}$ halos at $z=1.3-4$ is $\sim0.01$ $Z_{\odot}$, weakly dependent on halo mass and redshift. 
Similarly, \cite{2012MNRAS.424.2292G}, concluded that cold streams will be challenging to detect in metal line absorption due to their low-metallicity and small covering fractions. 
The low-metallicity cold streams found in simulations severely under-predict the metal line equivalent widths observed around LBGs, strongly suggesting that most of the metal absorption observed transverse to LBGs originates from gas that has been processed by galaxies, such as galactic winds. 
Indeed, \cite{2015MNRAS.450.2067T} used photoionization modeling to argue that at least some of the metal-enriched gas ($\gtrsim 0.1 Z_{\odot}$) observed around $z\sim2.3$ star-forming galaxies arises in the hot phase of galactic winds. 
In a few instances where abundance ratios have been measured, metal-rich CGM absorbers have abundance ratios consistent with either core collapse \citep[e.g.,][]{2015arXiv151006018L} or Type Ia supernovae \citep[e.g.,][]{2016MNRAS.458.2423Z}.
Cosmological simulations also convincingly demonstrate that star formation-driven galactic winds are necessary to explain metals observed in the CGM \citep[e.g.,][]{2012ApJ...760...50S, 2013MNRAS.430.1548H, 2015MNRAS.448..895S, 2016MNRAS.458.1164L, 2016MNRAS.459.1745F, 2016arXiv160508700T}. 

Despite the association between metal-rich gas and outflows, there is no clear cut metallicity division between inflows and outflows.  
In \S \ref{sec:HI_coverings}, we emphasized how galactic winds from infalling satellites can puff up large-scale structure filaments. 
Thus, a good fraction of the gas first accreting onto galaxies may come into contact with metal-enriched material. 
The extent to which this metal-enriched gas contaminates galactic accretion depends on how efficiently metals mix in the CGM. 
While simulations provide some indication of the expected mixing, observations of closely spaced sight lines toward gravitationally lensed quasars and photoionization modeling show that metal absorption systems are often compact and poorly mixed \citep[e.g.,][]{1999ApJ...515..500R, 2001ApJ...554..823R, 2006ApJ...637..648S, 2007MNRAS.379.1169S, 2015MNRAS.446...18C}. 
Overall, cool metal absorbers have inferred sizes ranging from $\sim 1$ pc to $\gtrsim 1$ kpc, with some evidence that typical size increases with increasing ionization state. 
Some clouds may be less than a solar mass in mass. 

All numerical methods are limited in their ability to capture metal mixing near their resolution limit. 
While grid codes tend to over-mix metals due to diffusive errors at the grid scale,\footnote{Such errors are mitigated in moving-mesh codes in which grid cells are advected with the flow, such as Arepo \citep[][]{2010MNRAS.401..791S}, as well as in the ``meshless finite mass'' (MFM) method implemented in GIZMO \citep[][]{2015MNRAS.450...53H}.} 
standard SPH codes ``lock'' metals into SPH particles. 
Because of this, it is often assumed that SPH under-mixes metals. 
However, this is only true for metal clumps above the resolution limit: tiny metal clumps below the resolution limit will appear over-mixed in SPH codes because their metals will be spread over the gas mass of individual SPH particles. 
For reference, state-of-the-art zoom-in SPH simulations of Milky Way-mass galaxies have typical gas particle masses $\sim10^{4}-10^{5}$ M$_{\odot}$ \citep[][]{2013ApJ...765...89S, 2013MNRAS.428..129S, 2014MNRAS.445..581H}. 
Thus, even state-of-the-art SPH simulations likely underestimate metal mixing in the CGM in at least some circumstances. 
In such circumstances, sub-resolution SPH models that attempt to model metal diffusion owing to unresolved turbulence \citep[e.g.,][]{2010MNRAS.407.1581S} could vitiate rather than improve the solution. 
An important question for future work will be to identify the kinds of CGM absorbers that can be reliably resolved in cosmological simulations. 
If metal absorbers are compact because supernova ejecta take a long time to mix with ambient gas, then absorption by hydrogen and helium (elements synthesized in the Big Bang) may not suffer from the same clumpiness effects. 
Warm and hot gas phases, which tend to be more volume-filling, may also be easier to resolve in simulations. 
OVI, now routinely detected at both low and high redshift \citep[e.g.,][]{2011Sci...334..948T, 2014ApJ...788..119L, 2015MNRAS.450.2067T}, stands out as one particular ion for which it will be important to determine the convergence properties of cosmological simulations.

Upcoming observations of multiple sight lines through the halos of individual galaxies will provide useful information regarding the size scales of different CGM absorbers. 
The best local laboratory to carry out such an experiment is M31. 
\cite{2015ApJ...804...79L} recently analyzed 18 sight lines within $2 R_{\rm vir}\approx 600$ kpc of M31, thus producing a partial map of the multi-phase CGM around the galaxy. 
New HST/COS observations of quasars behind M31 as part of the AMIGA (Absorption Maps In the Gas of Andromeda) project (PI N. Lehner) will improve on this pilot analysis with a total of 25 sight lines within $1.1R_{\rm vir}$. 
While these observations will not constrain structure on the fine scales possible with gravitationally lensed quasars or photoionization modeling, the spatially resolved map of M31's CGM will help us better interpret observations of single sight lines through the halos of similar-mass galaxies, such as those from the COS-Halos program \citep[][]{2011Sci...334..948T}. 
One caveat with drawing inferences based on detailed studies of a single system, however, is that simulations show that the CGM can be quite dynamic and time variable \citep[e.g.,][]{2015MNRAS.449..987F, 2016arXiv160805712H}. 
Thus, such observational analyses will be most powerful when combined with simulations that can inform how the observational inferences can be generalized to other halos. 
Another approach for observationally constraining the size scale of CGM structures is to quantify the fraction of the area of a background source absorbed by foreground CGM clouds. 
Quasar accretion disks have diameters $\sim0.01$ pc while galaxies generally have half-light radii $\gtrsim 1$ kpc. 
Thus, background galaxies probe absorber size scales larger than background quasars \citep[e.g.,][]{2016ApJ...824...24D}.

\subsection{Down-the-Barrel Metal Absorption Lines}
\label{sec:down-the-barrel}
Another observational approach to detect galactic accretion is to use single ``down-the-barrel'' spectra of galaxies \citep[e.g.,][]{2010ApJ...717..289S, 2012ApJ...747L..26R, 2012ApJ...760..127M}. 
One advantage of the down-the-barrel observations, relative to galaxy pairs, is that gas that absorbs stellar light from a galaxy is known to be located between the galaxy and the observer. 
Thus, redshifted absorption can be unambiguously associated with gas with a radial velocity component in the direction of the galaxy.\footnote{In galaxy pair experiments, an outflowing absorber located behind the foreground galaxy would also appear redshifted.  
This introduces a generic ambiguity in the interpretation of absorption lines transverse to foreground galaxies.} 
Down-the-barrel observations, however, suffer from a different difficulty due to the fact that the typical infall velocity of IGM accretion is comparable to the halo velocity.\footnote{Absent hydrodynamic interactions and angular momentum, the radial velocity would be simply equal to the velocity of free fall into halo. Current cosmological simulations indicate that asymptotic cold stream radial velocities are typically closer to half the halo circular velocity \citep[][]{2005MNRAS.363....2K, 2015MNRAS.450.3359G}.} 
Thus, even for sight lines fortuitously aligned with infalling CGM cool gas, absorption from the infalling gas will typically overlap in velocity space with interstellar medium (ISM) gas. 
Since ISM gas is expected to be generally both denser and more enriched with metals than cold streams, down-the-barrel absorption by cold streams will typically appear as a minor perturbation to ISM absorption \citep[][]{2011MNRAS.413L..51K}.\footnote{If the stellar light of a galaxy were concentrated in a point source and the ISM were rotating in perfectly circular motion around the center, then the ISM would move purely tangentially with respect to the light source and could not mimic infalling gas. In real galaxies, stellar light is however spatially extended and the internal dynamics and morphology of the ISM can be quite complex. For example, at $z\sim2$ the nebular line emission of galaxies is often very clumpy \citep[e.g.,][]{2009ApJ...706.1364F}. 
These effects could cause some ISM gas to appear as infalling.} 
In detailed analyses of down-the-barrel spectra, \cite{2012ApJ...747L..26R} and \cite{2012ApJ...760..127M} reported detections of infalling gas in a small fraction, $\sim 3-6\%$, of $z\sim0.4-1.4$ galaxies. 
These detections were made using low-ionization metal absorption lines and thus likely trace relatively metal-rich gas, such as infalling dwarf galaxies on their way to merging or recycling wind gas, rather than gas accreting from the IGM for the first time. 
Unfortunately, there has been relatively little modeling of the inflow signatures expected in down-the-barrel observations. 
Despite the challenges in using this technique for probing IGM accretion, more modeling would very valuable given the very rich observational datasets now available, which may make it possible to extract even subtle signatures. 

\begin{figure}[t]
\includegraphics[scale=.552]{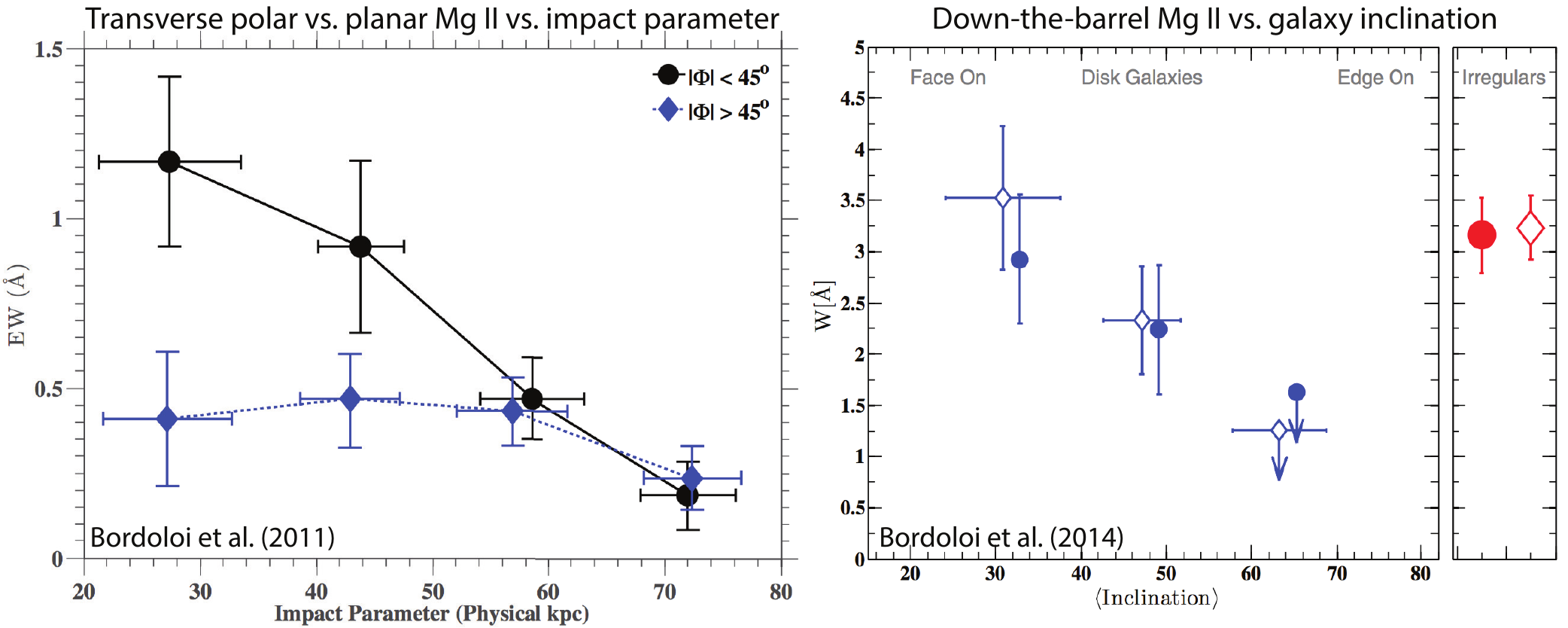}
\caption{Examples of how observed Mg II absorption varies as a function of azimuthal angle relative to the semi-minor axis of a galaxy and as a function of inclination angle in down-the-barrel spectra of galaxies. \emph{Left:} Average Mg II equivalent width around $0.5 \leq z \leq 0.9$ zCOSMOS disk galaxies as a function of impact parameter along the disk axis ($| \phi | < 45^{\circ}$) vs. near the disk plane \citep[$| \phi |>45^{\circ}$;][]{2011ApJ...743...10B}. \emph{Right:} Average MgII equivalent width vs. inclination from co-added down-the-barrel spectra of zCOSMOS $1\leq z \leq 1.5$ galaxies \citep[][]{2014ApJ...794..130B}. The circles and diamonds correspond to two different ways of making the measurement. 
These observations are consistent with galactic winds preferentially expanding normal to the plane of disk galaxies but there has been so far relatively little modeling of these observations.}
\label{fig:bordoloi}
\end{figure}

\subsection{Kinematic and Azimuthal Angle Diagnostics}
\label{sec:kinematics_angle}
A simple toy physical picture for inflows and outflows is one in which inflows from the IGM bring in the angular momentum that creates rotating galactic disks and in which galactic winds have a bi-conical morphology due to collimation normal to the galactic plane. 
If this toy model were correct, it would suggest that absorption by gas normal to the plane of disk galaxies should arise primarily from galactic winds, while absorption in the disk plane may be commonly due to infalling gas. 
In this picture, the infalling gas would typically co-rotate with the disk as it approaches the galaxy. 
Thus, a combination of azimuthal angle and kinematic diagnostics would constitute a powerful probe of inflows and outflows. 
This is indeed a promising avenue for identifying inflows and outflows, with some observational support for physical differences between planar and extra-planar absorbers in the CGM of galaxies. 
However, the latest high-resolution cosmological simulations indicate that the character of both galactic accretion and galactic winds change significantly with redshift and galaxy mass \citep[e.g.,][]{2015MNRAS.454.2691M, 2015arXiv151005650H}. 
Observations also show that galactic winds become significant weaker as star formation activity in galaxies declines from its peak at $z\sim2$ \citep[e.g.,][]{2010ApJ...717..289S} to the present \citep[e.g.,][]{2016ApJ...822....9H}. 
Thus, it is likely that the toy model outlined above is too simple, and this is an area where more detailed and systematic modeling is likely to prove critical. 
\begin{figure}[t]
\includegraphics[scale=0.575]{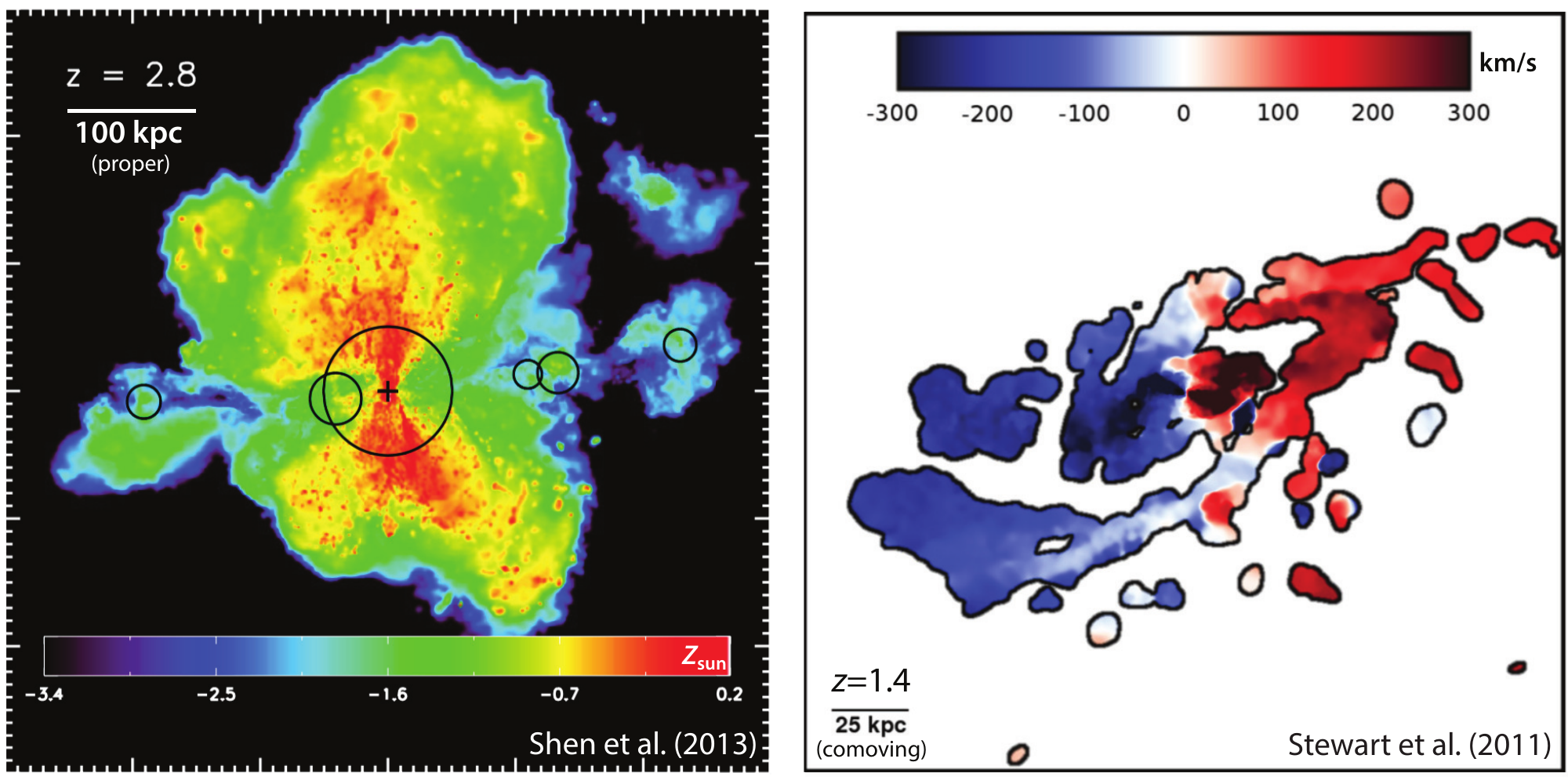}
\caption{Simulations that suggest the potential of azimuthal angle and kinematic diagnostics of galactic accretion. 
\emph{Left:} Projected gas metallicity in the Eris2 simulation at $z=2.8$. In this image, the stellar disk is nearly edge-on and the metal-enriched gas entrained by the galactic wind forms a rough bi-cone above and below the disk. 
The circles show virial radii, including of the five most massive nearby dwarf galaxies, which are aligned in the direction of metal-poor infalling filaments (adapted from Shen et al. 2013). 
\emph{Right:} Line-of-sight velocity map of cool halo gas ($N_{\rm HI}>10^{16}$ cm$^{-2}$) for a simulated Milky Way-mass halo at $z=1.4$. 
The cool halo gas tends to rotate in the same direction as cool halo gas out to $R\gtrsim40$ proper kpc (100 co-moving kpc). 
Both results shown here are based on zoom-in simulations of single halos, so more work is needed to determine which aspects generalize to different halo masses, assembly histories, redshifts, and details of how the baryonic physics is modeled. Adapted from Stewart et al. (2011).}
\label{fig:azimuthal_kinematic_sims}
\end{figure}

\subsubsection{Azimuthal Angle Diagnostics}
Observationally, there is support for galactic winds preferentially expanding normal to the plane of disk galaxies from spectroscopic observations transverse to foreground galaxies \citep[expressed in terms of azimuthal angle relative the semi-minor axis of the galaxy projected on the sky;][]{2011ApJ...743...10B, 2012MNRAS.426..801B, 2012ApJ...760L...7K} and from down-the-barrel spectra of galaxies as a function of inclination angle of the disk \citep[][]{2012ApJ...758..135K, 2012ApJ...747L..26R, 2014ApJ...794..156R, 2012ApJ...760..127M, 2014ApJ...794..130B}. 
Figure \ref{fig:bordoloi} shows observations of each type for $z\sim0.5-1.5$ galaxies in zCOSMOS. 
The larger average Mg II absorption equivalent widths along the poles of disk galaxies suggest that strong polar Mg II absorbers trace galactic winds. 
Because down-the-barrel sight lines toward low-inclination disk galaxies probe regions of the CGM similar to transverse spectra at small azimuthal (polar) angles, we will refer to both types of observations as probing the azimuthal angle dependence of CGM gas in what follows.\footnote{Note, however, that the two types of observations are not equivalent since down-the-barrel spectra are always sensitive to high-density material near (or within) the target galaxy, while transverse spectra only probe the CGM at distances from the galaxy equal to the impact parameter or greater.} 
In Mg II absorption, the observed azimuthal angle dependence appears to be stronger for systems with high rest equivalent widths $W_{0}\gtrsim$ 1~\AA. 
This is consistent with an origin of strong MgII absorbers in galactic winds, which is also supported by other observations \citep[e.g.,][]{2009MNRAS.393..808M, 2011MNRAS.412.1559N, 2012ApJ...761..112M}. 
Recently, \cite{2015ApJ...815...22K} reported evidence that OVI absorbers in the CGM of $0.08 \leq z \leq 0.67$ galaxies arise primarily either along their minor axis or their major axis, with stronger absorbers being preferentially found along the minor axis.  

Most existing observations of azimuthal angle dependence are at low to intermediate redshift, $z \lesssim 1.5$. 
This is in part because these observations require high-resolution imaging of the foreground galaxy in order to measure inclination or azimuthal angle. 
At high redshift, this requires long integrations with either HST or adaptive optics. 
In an analysis of the rest-frame optical morphological properties of $z\sim2-3$ star-forming galaxies, 
\cite{2012ApJ...759...29L} concluded that in contrast to galaxies at lower redshifts, there is no evidence for a correlation between outflow velocity and galaxy inclination. 
On the other hand, \cite{2012ApJ...761...43N} found 3$\sigma$ evidence in a sample of 27 $z\sim2$ star-forming galaxies with spatially resolved spectroscopic data that the mass loading factors of galactic winds are higher in face-on galaxies. 
It will be interesting to expand studies of azimuthal angle dependence in this redshift regime. 

From the theoretical standpoint, it is possible that any azimuthal dependence present at lower redshift will be either weaker or absent in the CGM of $z\gtrsim2$ galaxies. 
First, $z\gtrsim2$ star-forming galaxies are often observed to have clumpy morphologies, especially in the UV light that traces star formation \citep[e.g.,][]{2011ApJ...731...65F}, and galactic winds are to a large extent driven by outflows from prominent star-forming clumps \citep[e.g.,][]{2011ApJ...733..101G, 2016MNRAS.458.1891B}. 
Chaotic and clumpy galaxy morphologies at high redshift followed by the gradual emergence of stable disks are also a generic finding of recent cosmological simulations \citep[e.g.,][]{2009MNRAS.397L..64A, 2010MNRAS.404.2151C, 2014MNRAS.445..581H, 2016arXiv160303778O, 2016arXiv160804133M, 2016arXiv160802114C}. 
Thus, there may simply often not be well-defined gaseous disks to neatly collimate galactic winds at high redshift. 
Furthermore, at $z\sim2$ SFRs can be elevated relative to the local Universe by up to $\sim2$ orders of magnitude, and some simulations suggest that galactic wind bursts may be sufficiently powerful to expel most of the ISM from galaxies \citep[e.g.,][]{2015MNRAS.454.2691M}. 
In that case, even if it were present, a gaseous disk may not offer enough resistance to significantly collimate the galactic wind. 

Overall, azimuthal angle dependence is a promising approach for separating inflows and outflows statistically, but a more systematic analysis of the predictions of galaxy formation simulations will be needed to inform when a significant azimuthal angle dependence is expected. 
The left panel of Figure \ref{fig:azimuthal_kinematic_sims} shows an example of how the metallicity of CGM gas varies with azimuthal angle in the Eris2 simulation of a Milky Way progenitor at $z=2.8$ \citep[][]{2013ApJ...765...89S}. 
In this example, gas is substantially more metal enriched above and below the galaxy due to the effects of galactic winds, but it remains to be shown whether this simulation is representative. 

\subsubsection{Kinematic Diagnostics}
\label{sec:kinematic}
Weaker Mg II absorbers can arise from either inflows or outflows, but kinematics can potentially distinguish different origins. 
Several simulations indicate that accreting cool gas preferentially joins galactic disks in a co-rotating structure, consistent with galactic disks acquiring angular momentum from the accreting gas \cite[][]{2009ApJ...700L...1K, 2009ApJ...694..396B, 2011ApJ...738...39S, 2013ApJ...769...74S, 2016MNRAS.tmp.1114K, 2016arXiv160608542S}.\footnote{This is not to say that infalling cool gas solely determines the angular momentum of disk galaxies. In an analysis of the Illustris simulation, \cite{2016arXiv160801323Z} identify the important roles of specific angular momentum transfer from dark matter onto gas during mergers and from feedback expelling low specific angular momentum gas from halos.} 
Because the infalling gas co-rotates with the galaxy, its distribution is expected to be at least mildly flattened along the galactic semi-major axis. 
The right panel shows an example of a zoom-in simulation of a Milky Way-mass halo at $z=1.4$ in which the radial velocity profile of the cool halo  gas with HI column $N_{\rm HI}>10^{16}$ cm$^{-2}$ indicates co-rotation with the growing galactic disk out to $R\gtrsim40$ proper kpc \citep[][]{2011ApJ...738...39S}.

Observationally, there are several tentative detections of co-rotating MgII absorbers at low and high redshifts \citep[e.g.,][]{2002ApJ...570..526S, 2010ApJ...711..533K, 2013Sci...341...50B, 2016ApJ...820..121B}, but current observational samples are small and conflicting results have been reported \citep[e.g.,][]{2011ApJ...733..105K}. 
The simple picture of co-rotating MgII absorbers tracing galactic accretion is no doubt complicated by the different possible origins of MgII absorbers, including outflows, which can also carry angular momentum imparted as they are launched from a rotating disk. 
Over the next several years, kinematic diagnostics of halo gas will become increasingly interesting with the advent of a new generation of integral field surveys, including MaNGA \citep[][]{2015ApJ...798....7B}, KMOS$^{\rm 3D}$ \citep[][]{2015ApJ...799..209W}, and with MUSE \citep[e.g.,][]{2015A&A...575A..75B}, which will provide a new handle on the internal kinematics of galaxies. 
As with azimuthal angle diagnostics, systematically analyzing simulation predictions for the kinematics of gas galactic accretion relative to the orientation and internal kinematics of galaxies for a wide range of redshifts and galaxy properties will be critical to make progress.  
Currently, our theoretical expectations are limited by the small number of simulated halos for which kinematic relationships have been analyzed in detail, with existing studies being typically limited to a one or a few zoom-in simulations.

\begin{figure}[t]
\includegraphics[scale=.57]{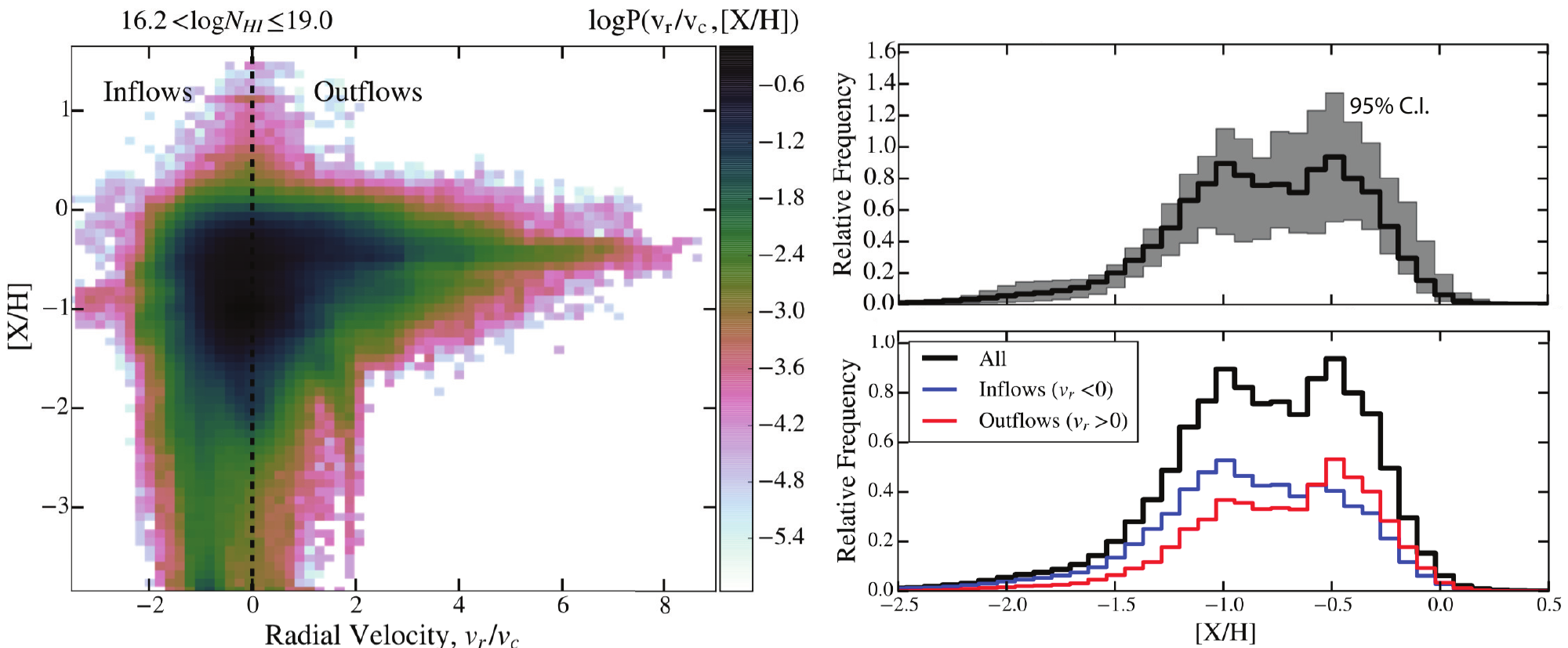}
\caption{Properties of the cosmological distribution of (randomly selected) LLSs at $0<z<1$ computed by convolving a suite of zoom-in simulations from the FIRE project with the dark matter halo mass function. 
\emph{Left:} Relative incidence (logarithmic units) of LLSs with $10^{16.2}< N_{\rm HI} \leq 10^{19}$ cm$^{-2}$ in the metallicity vs. radial kinematics plane (radial velocity $v_{\rm r}$ defined relative to the central galaxy of the halo hosting the LLS and expressed in units of halo circular velocity). 
High-velocity outflows (with radial velocity exceeding the circular velocity by a factor $\gtrsim2$) tend to have higher metallicities (${\rm [\rm X/H]}\sim -0.5$) while very low-metallicity LLSs (${\rm [X/H]}<-2$) are typically associated with gas infalling from the IGM. 
\emph{Right:} The corresponding overall LLS metallicity distribution. 
In the top panel, the gray region shows the 95\% confidence interval resulting from the limited number of zoom-in simulations included in the analysis. 
There is no significant evidence for multiple modes in the simulated metallicity distribution. 
In the bottom panel, the total metallicity distribution is divided between gas elements that are instantaneously inflowing ($v_{\rm r}<0$) and outflowing ($v_{\rm r}>0$) relative to their central galaxy. 
The inflowing and outflowing distributions overlap strongly in part because wind recycling is efficient at low redshift, so metal-enriched outflows are later identified as instantaneous inflows. 
Adapted from \cite{2016arXiv160805712H}. 
}
\label{fig:zach_lls}
\end{figure}

\subsection{Cosmological Absorber Statistics}
\label{sec:cosmlogical_stats}
All the observational diagnostics discussed so far rely on an association between absorbing gas and a galaxy. 
However, quasar absorption spectra contain a wealth of information on intergalactic absorbers without known galaxy associations. 
Nevertheless, many of the stronger absorption systems in the spectra of arbitrarily selected quasars arise in the CGM of foreground galaxies and thus provide important statistical constraints on galactic accretion. 

\subsubsection{The HI Column Density Distribution}
\label{sec:NHI_distribution}
We saw in \S \ref{sec:HI_coverings} that LLSs in galaxy halos arise from a mix of inflows and outflows. Thus, LLSs from galactic accretion contribute to the observed HI column density distribution. 
Using a simulation from the OWLS project post-processed with radiative transfer, \cite{2012MNRAS.421.2809V} quantified the contribution of cold accretion flows to the observed $z=3$ HI column density distribution. 
The simulation analyzed by \cite{2012MNRAS.421.2809V} reproduces the observed HI column density distribution over ten orders of magnitude in $N_{\rm HI}$ \citep[][]{2011ApJ...737L..37A}.  
In this simulation most LLSs arise within galaxy halos and most of these are infalling toward a nearby galaxy. 
On this basis, \cite{2012MNRAS.421.2809V} concluded that cold accretion flows predicted by cosmological simulations have been statistically detected in the observed HI column density distribution at $z=3$. 
The argument is compelling, though there are some caveats that future simulations and observational analyses should attempt to address to firm up the conclusion. 
Observationally, measurements of the column density distribution are relatively uncertain in the LLS regime \citep[e.g.,][]{2010ApJ...718..392P}, in part because LLSs are on the flat part of the curve of growth. 
Theoretically, the simulation results summarized in \S \ref{sec:HI_coverings} indicate that galactic winds can contribute comparably to -- or even dominate over -- cold accretion streams in explaining LLSs in galaxy halos at $z\sim2-2.5$. 
Since the properties of galactic winds are uncertain, it is plausible that reasonable agreement with the observed column density distribution could be obtained absent cold streams for some wind models. 
Finally, a combination of resolution effects and numerical limitations of different hydrodynamic solvers introduces additional uncertainties in the theoretical predictions \citep[][]{2013MNRAS.429.3341B, 2013MNRAS.429.3353N, 2016MNRAS.460.2881N}.

\subsubsection{The Metallicity Distribution of LLSs}
\label{sec:LLS_metal_distr}
Recently, \cite{2013ApJ...770..138L} and \cite{2016arXiv160802584W} reported evidence that the metallicity distribution of randomly selected LLSs at $z<1$ is bimodal, with dearth of LLSs with metallicity of about ten percent solar.\footnote{In the updated $z \leq 1$ metallicity analysis of \cite{2016arXiv160802584W}, the statistical evidence for a bimodality is strongest for a subsample restricted to partial LLSs, with $16.2 \leq \log{N_{\rm HI}} \leq 17.2$.} 
These authors interpreted the high-metallicity branch as arising in outflows, recycling winds, and tidally stripped gas around galaxies, while the low-metallicity branch may trace cool, dense accreting gas. 
If this interpretation is correct, then LLS metallicity would be an extremely powerful way to identify cool galactic accretion at $z<1$. 
At $z>2$, the observational analyses of \cite{2016MNRAS.455.4100F}, \cite{2016arXiv160802588L}, and \cite{2016arXiv160402144G} indicate instead a broad unimodal distribution of LLS metallicities. 

Motivated by the low-redshift observations of \cite{2013ApJ...770..138L} and \cite{2016arXiv160802584W}, \cite{2016arXiv160805712H} analyzed the LLS metallicity distribution at $z<1$ using a sample of zoom-in simulations from the FIRE project. 
To model the cosmological distribution from a sample of zoom-in simulations, \cite{2016arXiv160805712H} convolved the LLS properties for individual halos with the dark matter halo mass function. 
In these simulations, LLSs are concentrated close to galaxies at $z<1$ so this halo-based approach should capture the majority of LLSs; \cite{2016arXiv160805712H} showed that it reproduces the LLS cosmological incidence measured by \cite{2011ApJ...736...42R}. 
Figure \ref{fig:zach_lls} summarizes the key results from \cite{2016arXiv160805712H} regarding the LLS metallicity distribution, and the relationship of LLS metallicity with inflows and outflows defined based on instantaneous radial kinematic relative to central galaxies. 
The analysis indicates that high-velocity outflows (with radial velocity exceeding the halo circular velocity by a factor $\gtrsim2$) tend to have higher metallicities (${\rm [X/H]}\sim -0.5$) while very low-metallicity LLSs  (${\rm [X/H]}<-2$) are typically associated with IGM inflows. 
However, most LLSs occupy an intermediate region in metallicity-radial velocity space. 
Overall, the simulated LLS metallicity distribution does not show significant evidence for bimodality.
The strong overlap between instantaneous inflows and outflows for intermediate metallicity systems is in part due to the prevalence of wind recycling in the FIRE simulations at $z<1$, which causes metal-rich galactic wind ejecta to later fall back onto galaxies~\citep{Angles-16}.
The lack of a clean bimodality in the simulated LLS metallicity distribution is also due to the fact that halos covering the broad mass range $M_{\rm h}\sim10^{10} - 10^{12}$ M$_{\odot}$ contribute significantly to the distribution. 
Since the ISM and CGM metallicities both increase with galaxy mass in the simulations \citep[][]{2016MNRAS.456.2140M, 2016arXiv160609252M}, any narrow feature in the metallicity distribution is likely to be washed out in the cosmological average. 
One effect that could cause simulations to miss features in the metallicity distribution is the mixing of metals on small scales. 
As mentioned above, some observations indicate that metals can be locked in compact clumps that will not be resolvable in cosmological simulations for the foreseeable future \citep[e.g.,][]{2007MNRAS.379.1169S}. 
Going forward, it will be useful to address this issue by supplementing cosmological simulations with higher-resolution calculations better suited to understand small-scale mixing. 
The analysis of \cite{2016arXiv160805712H} included only 14 simulated main halos, so it will also be important to firm up the statistical significance of the results. 
Furthermore, it will be interesting to use simulations to study in more detail the redshift evolution of LLS metallicities, as well as how the metallicity distribution changes with HI column, e.g. from LLSs to DLAs. 
\cite{2015ApJ...812...58C} analyzed the $z=3.5$ LLS metallicity distribution in a full-volume cosmological simulation with the Illustris galaxy formation model \citep[][]{2013MNRAS.436.3031V, 2014MNRAS.438.1985T} and also found a broad unimodal distribution.

\section{Emission Diagnostics}
\label{sec:emission}
CGM emission is a probe of galactic accretion complementary to absorption measurements. 
The principal advantage of emission measurements is that they provide spatial resolution in individual halos, which can be used to identify galactic accretion using morphological signatures, such as filaments. 
At present, the main challenge with emission observations is that circum-galactic gas is typically very faint, so emission measurements are currently only possible for dense gas relatively close to galaxies, or in halos with a luminous quasar that can power CGM emission out to larger radii. 
High-quality CGM emission observations will, however, become increasingly common over the next few years as a number of optical integral field spectrographs (IFS) with the capacity to detect low surface brightness, redshifted rest-UV CGM emission have recently been commissioned or are planned for the near future. 
The Palomar Cosmic Web Imager \citep[PCWI;][]{2010SPIE.7735E..24M} started taking data in 2009 and the first science results on luminous spatially extended Ly$\alpha$ sources at $z\sim2-3$ have been reported \citep[][]{2014ApJ...786..106M, 2014ApJ...786..107M}. 
Its successor, the Keck Cosmic Web Imager \citep[KCWI,][]{2010SPIE.7735E..21M}, to be mounted on the Keck II telescope, is currently being developed. 
The Multi-Unit Spectroscopic Explorer \citep[MUSE;][]{2010SPIE.7735E...7B} on the Very Large Telescope (VLT) completed its commissioning in August 2014 and early science results are being reported \citep[e.g.,][]{2016MNRAS.462.1978F, 2016A&A...587A..98W, 2016arXiv160501422B}. 
These IFSs provide kinematic information not available with narrowband imaging, and their spectroscopic resolution also enables more accurate background subtraction for line emission. 

In this section, we provide a brief summary of simulation results regarding Ly$\alpha$ emission (\S \ref{sec:lya_em}), UV metal line emission (\S \ref{sec:meta_emission}), and X-ray emission from the CGM (\S \ref{sec:hot_gas}), as well as a summary of the observational status for each.

\begin{figure}[t]
\begin{center}
\includegraphics[scale=.65]{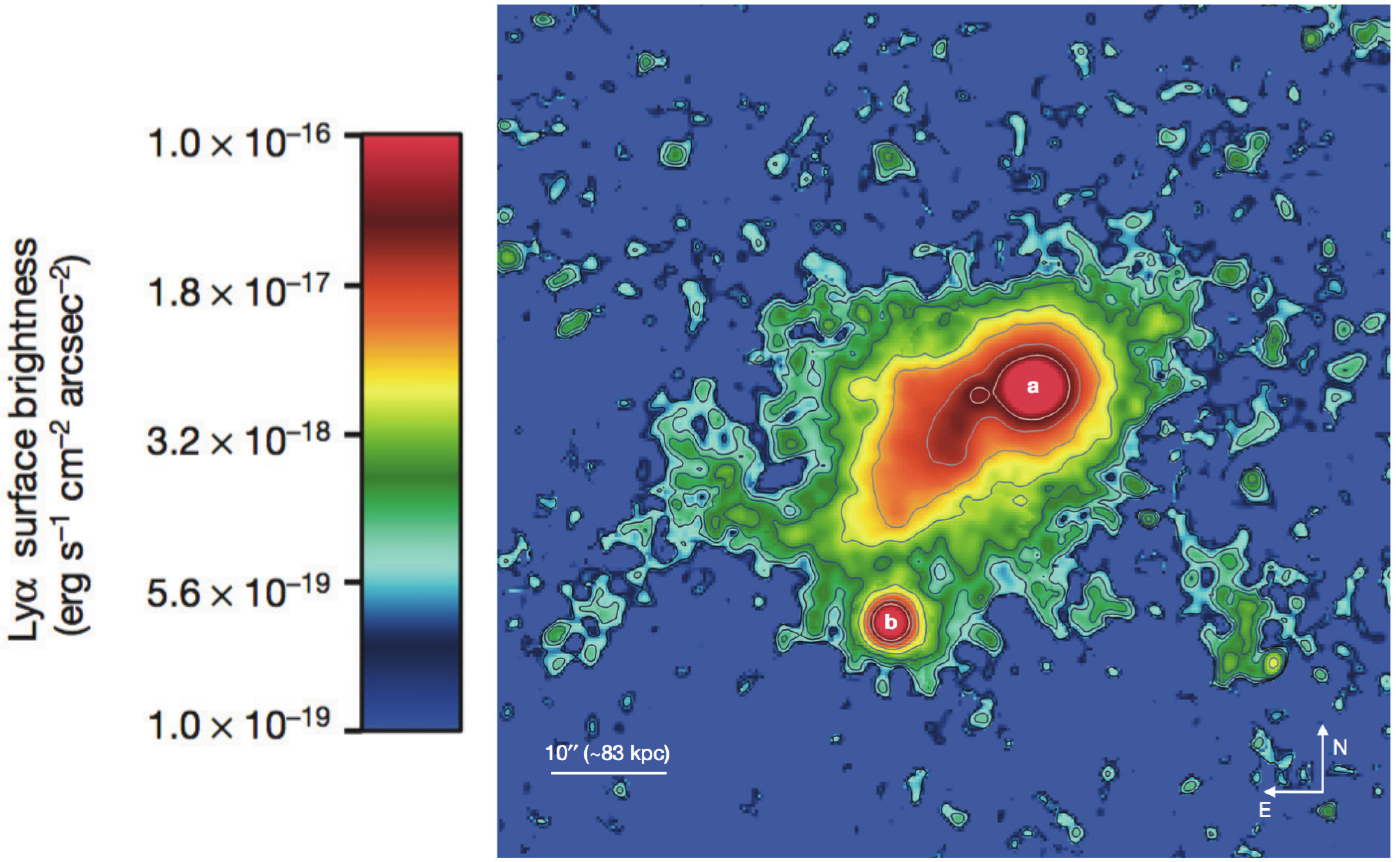}
\end{center}
\caption{
Ly$\alpha$ image of the nebula around the UM 287 quasar (`a') at $z\approx2.3$ (adapted from Cantalupo et al. 2014). 
The second bright spot labeled `b' marks the location of another, optically faint quasar at the same redshift. 
The extended filamentary morphology suggests that the Ly$\alpha$ emission traces a cold accretion flow. 
Follow-up integral field observations suggest a smooth kinematic profile consistent with a giant, rotating proto-galactic disk for the brightest portion of the filament, which appears to connect smoothly to the cosmic web \citep[][]{2015Natur.524..192M}. 
}
\label{fig:cantalupo}
\end{figure}

\subsection{Ly$\alpha$ Emission from the CGM}
\label{sec:lya_em}
Ly$\alpha$ emission is typically the brightest emission line from the CGM. 
Our first glimpses of CGM emission have indeed come from spatially extended Ly$\alpha$ sources known as ``Ly$\alpha$ blobs'' (LABs).  
The classical LABs have line luminosities up to $\sim 10^{44}$ erg s$^{-1}$ and spatial extents sometimes exceeding $100$ proper kpc \citep[][]{2000ApJ...532..170S, 2004AJ....128..569M, 2009ApJ...693.1579Y}. 
The physical nature of LABs is not yet well understood, but several studies suggested that they could be powered by the conversion of gravitational potential into Ly$\alpha$ photons as gas accretes onto halos or galaxies (``cooling radiation''). 
In this model, weak shocks continuously heat cold accreting gas to temperatures $T\sim10^{4}$ K and this energy is efficiently converted into Ly$\alpha$ emission via collisional excitation of HI \citep[][]{2000ApJ...537L...5H, 2001ApJ...562..605F, 2009MNRAS.400.1109D, 2010MNRAS.407..613G, 2012MNRAS.423..344R}. 
However, a major hurdle in identifying diffuse Ly$\alpha$ radiation with this process is that the expected luminosity remains uncertain at the order-of-magnitude level. 
Fundamentally, this is because the Ly$\alpha$ collisional excitation emissivity is an exponentially steep function of temperature near $T=10^{4}$ K, so that small errors in the thermodynamic history of the accreting cold gas can result in large differences in the predicted Ly$\alpha$ luminosity \citep[e.g.,][]{2010ApJ...725..633F}. 
There are two sources of error that can affect the thermodynamic history of accreting gas in cosmological simulations. 

The first is the accuracy of the hydrodynamics, which must be able to correctly capture the properties of both weak \emph{and} strong shocks experienced by accreting gas.  
The latter point regarding strong shocks is also important for the identification of cold accretion flows in simulations, and is worth expanding on. 
In both particle-based and grid-based hydrodynamic codes, shocks are often broadened across several resolution elements, which can lead to ``in-shock cooling'' \citep[e.g.,][]{2000MNRAS.319..721H, 2011MNRAS.415.3706C}. 
This problem arises, for example, when a strong shock should produce hot $T\gtrsim10^{6}$ K gas with a long cooling time but in the code the gas cools artificially as it passes through the numerically broadened shock and encounters the peak of the cooling function at $T\sim10^{5}$ K. 
In such circumstances, the hydrodynamic solver can overestimate radiative energy losses via low-energy processes, including Ly$\alpha$. 
A specific situation where this effect likely occurs in cosmological simulations is when cool accreting gas impacts a galaxy, where cooling times can be very short owing to the relatively high local gas densities. 
In this case, not only will there be an error in the predicted Ly$\alpha$ emission, but also artifacts can be introduced in simple algorithms for identifying cold mode accretion based on the maximum temperature to which gas is heated \citep[e.g.,][]{2005MNRAS.363....2K, 2009MNRAS.395..160K, 2011MNRAS.414.2458V, 2013MNRAS.429.3353N}. 

The second reason for the large uncertainties in predicted Ly$\alpha$ cooling luminosities is the treatment of ionizing radiative transfer. 
As discussed in \S \ref{sec:HI_coverings}, cold accretion streams are traced by LLSs, which by definition are optically thick to ionizing photons at the Lyman edge. 
Since most cosmological simulations to date do not include self-consistent ionizing radiative transfer, they do not accurately capture photoheating in dense self-shielded gas \citep[but see][]{2012MNRAS.423..344R}. 
In particular, simulations run with a uniform cosmic UVB and assuming optically thin ionization balance overestimate the amount of photoheating within cold streams. 
\cite{2010ApJ...725..633F} tested the sensitivity of their predictions to the treatment of dense gas and found that different assumptions produced Ly$\alpha$ luminosities differing by up to $\sim2$ orders of magnitude. 

Even if a significant fraction of the Ly$\alpha$ emission in LABs comes from cooling radiation, it is difficult to observationally separate cooling radiation from Ly$\alpha$ photons produced by other processes, such as star formation or AGN activity in galaxies. 
One reason is that bright cooling radiation requires high galactic accretion rates, which lead to SFRs (or AGN activity) sufficient to power most observed diffuse Ly$\alpha$ halos. 
Diffuse Ly$\alpha$ halos are now in fact generically observed around ordinary star-forming galaxies \citep[e.g.,][]{2011ApJ...736..160S} and these observations are consistent with the Ly$\alpha$ photons being produced by star formation inside galaxies. 
There are several ways in which star formation or AGN power can mimic the spatially extended emission expected from galactic accretion. 
Ly$\alpha$ photons produced inside galaxies can result in diffuse halos due to scattering of the Ly$\alpha$ photons in the CGM \citep[e.g.,][]{2012MNRAS.424.1672D}. 
Ionizing photons that escape galaxies but are absorbed in the CGM can also produce fluorescent Ly$\alpha$ emission \citep[e.g.,][]{1996ApJ...468..462G, 2005ApJ...628...61C, 2010ApJ...708.1048K}. 
Alternatively, energy injected in the CGM as galactic winds driven by stellar or AGN feedback encounter halo gas can also power diffuse Ly$\alpha$ emission \citep[][]{2000ApJ...532L..13T, 2001ApJ...562L..15T}. 
Since Ly$\alpha$ photons typically scatter a large number of times before escaping the CGM, the apparent Ly$\alpha$ spatial and velocity extents are not necessarily representative of the gas producing the Ly$\alpha$ photons.

A more promising avenue for using Ly$\alpha$ emission to identify galactic accretion is to simply use the Ly$\alpha$ photons as a tracers of CGM gas at last scattering. 
For example, many Ly$\alpha$ sources have a filamentary morphology reminiscent of cosmic web filaments and their extensions into galactic halos as cold streams \citep[e.g.,][]{2011MNRAS.418.1115R, 2014Natur.506...63C, 2014ApJ...786..106M, 2014ApJ...786..107M, 2011MNRAS.418.1115R, 2013MNRAS.429..429R, 2016MNRAS.455.3991R}. 
Of course, one must be careful not attribute every filamentary features to a cold accretion stream, since other phenomena such as tidally stripped gas can appear elongated. 
Nevertheless, a statistical study of the morphological properties of spatially extended Ly$\alpha$, along with a comparison to the incidence rate of accreting filaments predicted by cosmological simulations, can test simulation predictions for galactic accretion.  
In at least one case with detailed spatially resolved kinematic observations (the Ly$\alpha$ image shown in Fig. \ref{fig:cantalupo}), there is evidence that the filamentary structure traced by Ly$\alpha$ emission smoothly connects to a large, rotating proto-galactic disk \citep[][]{2015Natur.524..192M}.

Observations of particularly extended and luminous Ly$\alpha$ nebulae at high redshift provide further evidence for compact dense clumps in the gaseous halos of massive galaxies (see \S \ref{sec:metal_abs} for evidence from absorption measurements). 
Even if a luminous quasar can in principle power the observed Ly$\alpha$ luminosity through reprocessing of its ionizing radiation in the CGM, the integrated recombination rate over the nebula ($\propto \int dV \alpha(T) n_{\rm e} n_{\rm HII}$, where $\alpha$ is the hydrogen recombination coefficient) must be sufficiently high to account for the Ly$\alpha$ luminosity attributed to fluorescence. 
Recent detailed analyses of giant Ly$\alpha$ nebulae indicate that in at least some cases the Ly$\alpha$-emitting gas must be highly clumped and reach densities $\sim1$ cm$^{-3}$ (more typical of ISM gas than CGM gas) over spatial scales $\sim 100$ kpc \citep[e.g.,][]{2014Natur.506...63C, 2015Sci...348..779H}. 
If giant proto-galactic disks are relatively common at high redshift, one possibility is that much of the observed Ly$\alpha$ luminosity in giant nebulae originates in fluorescence due to a luminous quasar shining on such a disk rather than CGM gas \citep[e.g.,][]{2015Natur.524..192M}. 

The contribution by S. Cantalupo in this volume provides a more exhaustive review of recent results on spatially extended Ly$\alpha$ sources. 

\begin{figure}
\sidecaption
\includegraphics[scale=.66]{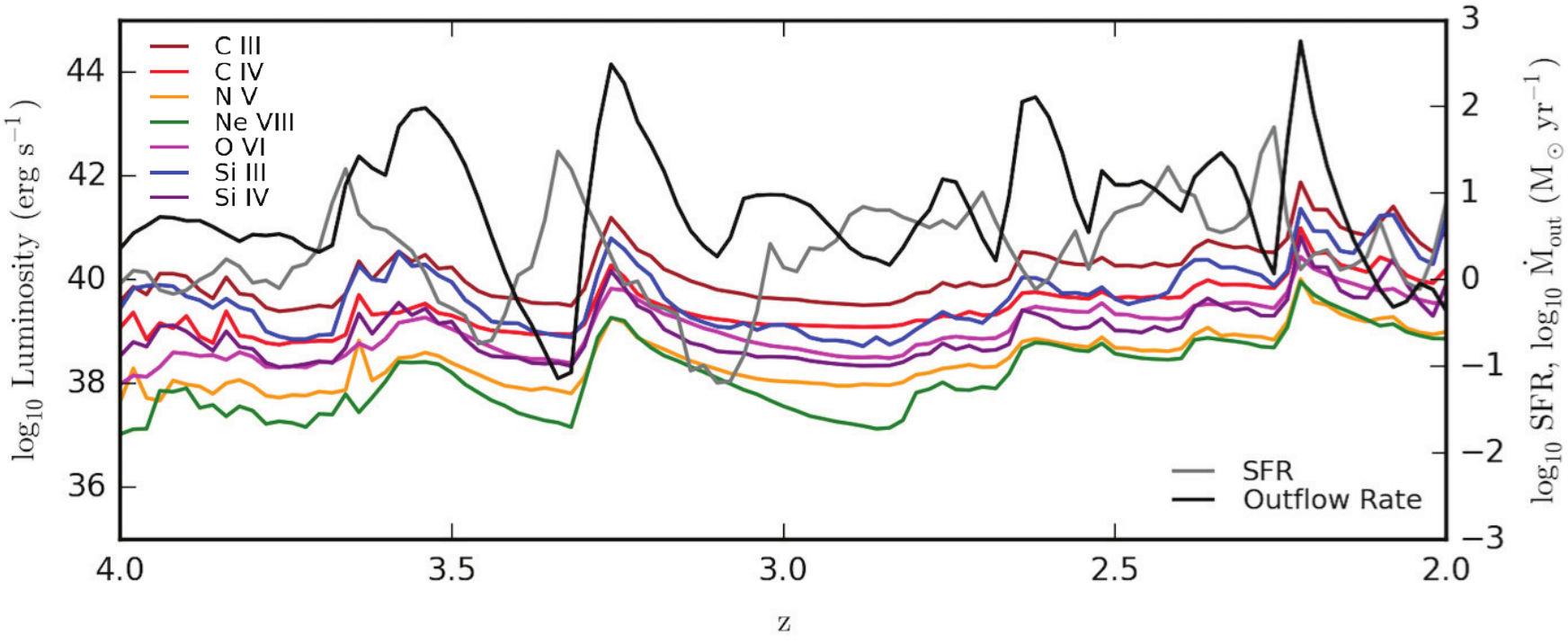}
\caption{Example simulation of the UV metal line emission from the CGM of an LBG-mass halo as a function of redshift. 
This simulation, from the FIRE project, includes strong stellar feedback. 
Colored lines show UV metal line luminosities within $1~R_{\rm vir}$ but excluding the inner 10 proper kpc (a proxy for central galaxies). 
Star formation rates within $1~R_{\rm vir}$ and gas mass outflow rates at $0.25~R_{\rm vir}$ are plotted as gray and black lines, respectively. 
The UV metal line luminosities, star formation, and mass outflow rates are all strongly time variable and correlated. 
Peaks in CGM luminosity correspond more closely with peaks in mass outflow rates, which typically follow peaks of star formation with a time delay, indicating that energy injected by galactic winds is the primary source of CGM UV metal line emission. 
Adapted from \cite{2015arXiv151006410S}.}
\label{fig:sravan_tdep}
\end{figure}

\subsection{UV Metal Line Emission from the CGM}
\label{sec:meta_emission}
Because metals are not as abundant, metal line emission is generally significantly fainter than Ly$\alpha$.  Metal lines can, however, provide very useful complementary information on the physical conditions in the CGM. 
Since most metal lines are optically thin, they are not subject to photon scattering effects and therefore more directly probe the spatial distribution and kinematics of the emitting gas. 
Furthermore, different metal ions probe different temperature regimes \citep[e.g.,][]{2012MNRAS.420.1731F, 2013MNRAS.430.2688V, 2016arXiv160708616C}. 
On the other hand, since metal emission preferentially probes metal-enriched gas, it is at present typically more useful as a general probe of the conditions in the CGM rather than of galactic accretion directly. 
For example, in an analysis of the UV metal line emission from the CGM of $z=2-4$ simulated LBGs from the FIRE project, \cite{2015arXiv151006410S} showed the UV metal line emission arises primarily from gas collisionally excited by galactic winds (see Fig. \ref{fig:sravan_tdep}). 

\begin{figure}[t]
\includegraphics[scale=.575]{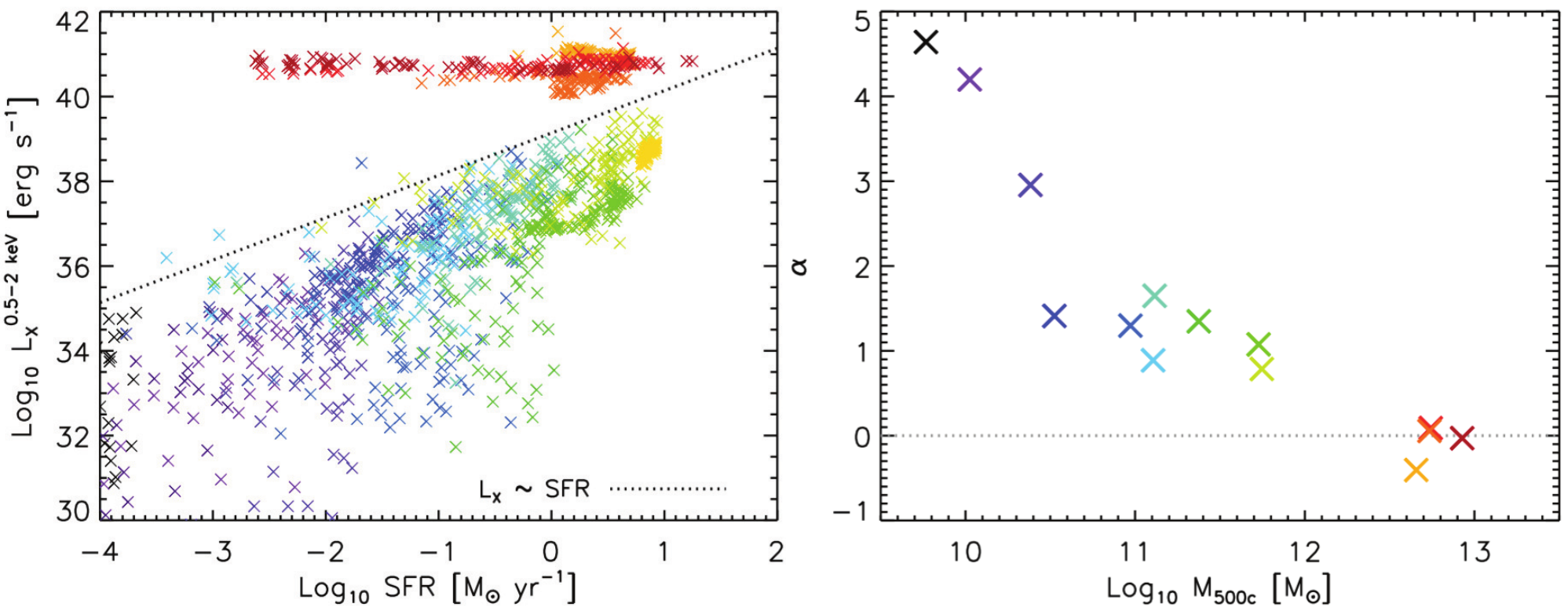}
\caption{
\emph{Left:} Soft X-ray luminosity $L_{\rm X}$ (0.5-2 keV) at $z=0-0.5$ as a function of SFR (averaged over 100 Myr) for zoom-in cosmological simulations with stellar feedback from the FIRE project. 
Crosses with the same color belong to the same galaxy at different times (halo masses can be read off from the panel on the right). 
\emph{Right:} The power, $\alpha$, of the correlation between $L_{\rm X}$ and SFR ($L_{\rm X} \propto SFR^{\alpha}$) for individual halos as a function of halo mass. 
The X-ray emission around dwarf galaxies is a strong function of their SFR, while halos with $M_{\rm 500c} \approx 10^{11-12}$ M$_{\odot}$ exhibit a correlation close to linear. 
There is no correlation between $L_{\rm X}$ and SFR for the most massive halos, because hot, virialized halo gas produces more X-rays than star formation-powered winds in those halos.  
Thus, X-ray emission is sensitive to gas accretion onto non-dwarf halos at low redshift (including Milky Way-mass halos, galaxy groups, and galaxy clusters) but primarily probes galactic winds in dwarfs. 
Adapted from \cite{2016arXiv160401397V}. 
}
\label{fig:van_de_voort}
\end{figure}

\subsection{X-ray Emission from Hot Halo Gas}
\label{sec:hot_gas}
Finally, we comment on the use of X-ray observations for probing galactic accretion. 
In galaxy clusters, it is well established that the hot intra-cluster medium (ICM) is primarily the result of gas shocked heated during the cluster assembly and that the ICM cooling observed in X-rays drives accretion onto galaxies \citep[albeit with a strong apparent suppression of the cooling flows in many clusters, tentatively due to AGN feedback, e.g.;][]{2007ARA&A..45..117M}. 
But what processes do X-rays probe in lower-mass halos \citep[e.g.,][]{2010ApJ...715L...1M, 2013ApJ...762..106A, 2016arXiv160802033L}? 

\cite{2016arXiv160401397V} analyzed the X-ray emission in simulated halos from the FIRE project. 
As for the other FIRE simulations mentioned in this review, these simulations included stellar feedback but no AGN feedback.  
Figure \ref{fig:van_de_voort} summarizes summarizes how the soft X-ray emission depends on SFR at $z<0.5$, for different halo masses. 
The X-ray emission around dwarf galaxies is a strong function of their SFR but there is no correlation between $L_{\rm X}$ around massive galaxies or galaxy groups ($M_{\rm 500c}>10^{12}$ M$_{\odot}$). 
In intermediate-mass halos ($M_{\rm 500c} \approx 10^{11-12}$ M$_{\odot}$), there is a close to linear relation between X-ray luminosity and SFR. 
These results indicate that diffuse X-rays primarily probe star formation-driven galactic winds in low-mass halos \citep[see also the analytic wind models of][]{2014ApJ...784...93Z}. 

\section{Conclusions and Outlook}
\label{sec:conclusions}
A common thread of this review is that there is no silver bullet in the quest to test models of galactic accretion. 
Each observational diagnostic that we discussed has not only some advantages, but also some ambiguities. 
There is in general likely no simple criterion that can be used to robustly identify galactic accretion in individual measurements. 
This is because the CGM is a complex and dynamic environment in which galactic accretion interacts with galactic winds, satellite galaxies, as well as more quiescent ambient gas. 
Furthermore, simulations predict that -- like galaxies - the properties of the CGM evolve significantly with redshift and halo mass. 
This complexity underscores the crucial role that simulations will continue to play in testing models of galactic accretion and feedback. 
Indeed, our current understanding points toward systematic, statistical comparisons with full-physics cosmological simulations as a necessary step to test the models. 

We conclude with a brief list of general areas in which progress is likely to be particularly fruitful over the next few years:
\begin{enumerate} 
\item Since cosmological simulations cannot resolve some of the fine-scale structure apparent in CGM observations (see \S \ref{sec:metal_abs}), it will be important to clarify which observational tracers can be robustly compared with simulation predictions. For example, are the main observable characteristics of the more massive and volume-filling CGM phases reasonably converged?
\item Relatedly, a better understanding of how metals returned by stellar evolution mix with ambient gas (both inside galaxies and after being ejected into the CGM) will ultimately be essential to make robust predictions for the metallicity distribution of CGM gas. 
Current sub-grid models for metal mixing due to unresolved turbulence rely on simplified schemes and do not account for the fact that metals may be injected as compact clumps well below the resolution limit. 
\item Non-ideal hydrodynamic effects, such as magnetic fields and thermal conduction, affect the survival and phase structure of CGM clouds, and should therefore be investigated. 
\item With the exception of a few recent studies, most previous simulations used to study diagnostics of galactic accretion either neglected galactic winds or used stellar feedback too weak to reproduce observed outflows and galaxy stellar masses. 
To develop reliable accretion diagnostics, it is critical to use feedback models that reproduce observed galaxy properties. 
\item Some promising diagnostics of inflows and outflows (e.g., azimuthal angle and kinematic diagnostics; \S \ref{sec:kinematics_angle}) have so far only been studied using small samples of simulated halos and for limited redshift ranges. 
Since galaxies and their CGM evolve strongly with redshift and mass, it will be necessary to analyze larger simulation samples that systematically cover relevant mass and redshift ranges to quantify the statistical robustness and limitations of the diagnostics.
\item Given the ambiguities of different inflow/outflow diagnostics when applied in isolation, quantifying how different diagnostics (e.g., metallicity, azimuthal angle, and kinematics) could be \emph{jointly} used to distinguish between inflows and outflows would be very useful.
\item Differential studies comparing observations at epochs where inflows/outflows are predicted to be more/less prominent could also help in breaking degeneracies in single observations.
\end{enumerate}
We note that most of these issues are not specific to the CGM, but of general importance to galaxy formation. 
It is thus clear that studies of the CGM will remain a very active area at the forefront of research on galaxy evolution for the foreseeable future.

\begin{acknowledgement}
We are grateful to many colleagues and collaborators who have helped shape our views on galactic accretion, including: Chuck Steidel, Gwen Rudie, Alice Shapley, Xavier Prochaska, Joe Hennawi, Michele Fumagalli, Nicolas Lehner, Chris Howk, Lars Hernquist, Joop Schaye, Freeke van de Voort, Andrey Kravtsov, Cameron Liang, Mark Dijkstra, Norm Murray, Eliot Quataert, Dusan Kere\v{s}, Phil Hopkins, Alexander Muratov, Daniel Angl\'es-Alc\'azar, and Zach Hafen. 
Our research on galactic accretion has been supported by NSF and NASA. 

\end{acknowledgement}


\end{document}